\documentclass[12pt, a4paper]{article}

\usepackage[utf8]{inputenc}
\usepackage[T1]{fontenc}
\usepackage{geometry}
\usepackage{setspace} 
\usepackage{amsmath, amssymb, amsfonts} 
\usepackage{graphicx} 
\usepackage{float}    

\usepackage{booktabs} 
\usepackage{array}    
\usepackage{tabularx} 
\usepackage{enumitem} 
\usepackage{titlesec} 

\usepackage[authoryear]{natbib} 

\usepackage{hyperref} 
\hypersetup{
    colorlinks=true,
    linkcolor=blue,
    filecolor=magenta,      
    urlcolor=cyan,
    citecolor=blue,
}

\title{\textbf{The Fractured Metropolis: Optimization Cutoffs, Uneven Congestion, and the Spatial Politics of Globalization}}
\author{Dong Yang}
\date{\today}

\begin{document}

\maketitle

\begin{abstract}
The divergence in globalization strategies between the US (retrenchment and polarization) and China (expansion) presents a puzzle that traditional distributional theories fail to fully explain. This paper offers a novel framework by conceptualizing the globalized economy as a ``Congestible Club Good,'' leading to a ``Fractured Metropolis.'' We argue that globalization flows ($M$) are constrained by domestic Institutional Capacity ($K$), which is heterogeneous and historically contingent.

We introduce the concept of the \textbf{``Optimization Cutoff''}: globalization incentivized the US to bypass costly domestic upgrades in favor of global expansion, leading to the long-term neglect of Public Capacity ($K_{Public}$). This historical path created a deep polarization. ``Congested Incumbents,'' reliant on the stagnant $K_{Public}$, experience globalization as chaos ($MC > MB$), while ``Insulated Elites'' use Private Capacity ($K_{Private}$) to bypass bottlenecks ($MB > MC$). This divergence paralyzes the consensus needed to restore $K_{Public}$, creating a \textbf{``Capacity Trap''} where protectionism becomes the politically rational, yet economically suboptimal, equilibrium.

Empirically, we construct an Institutional Congestion Index using textual analysis (2000-2024), revealing an exponential surge in disorder-related keywords (from 272 hits to 1,333). We triangulate this perception with the material failure of $K_{Public}$, such as the 3.7 million case backlog in US immigration courts. Our findings suggest the crisis of globalization is fundamentally a crisis of uneven institutional capacity and the resulting political paralysis.
\end{abstract}

\vspace{0.5cm}
\noindent \textbf{Keywords:} Globalization; Congestion Cost; Optimization Cutoff; Institutional Capacity; Polarization; US-China Relations; Political Economy

\newpage

\section{Introduction}

\subsection{A Tale of Two Globalizations}
In a congested metropolis, infrastructure dictates destiny. While the majority struggles with overwhelmed public transit, crowded schools, and backlogged administrative services, a privileged minority bypasses these bottlenecks entirely. Through private transportation, elite education, and exclusive networks, they access a parallel infrastructure that allows them to navigate the city efficiently.

This image of a fractured metropolis is not merely an urban phenomenon; it captures the new political economy of globalization within the United States. The past decade has witnessed a dramatic ``Great Reversal.'' The United States, once the architect of the liberal international order, has become its leading skeptic. While this shift is well-documented, the deeper puzzle lies in the profound internal polarization accompanying it.

Globalization is simultaneously hailed as an engine of unprecedented prosperity by some, and condemned by others as a source of systemic chaos—characterized by administrative paralysis, logistical breakdowns, and social friction. This paper argues that this polarization is not merely a disagreement over economic benefits, but a collision of two fundamentally different realities.

The crisis of globalization is a crisis of \textit{capacity}. The American political and physical infrastructure is overwhelmed by the flows it once championed. Crucially, the resulting congestion is experienced heterogeneously. The political battle over globalization is, fundamentally, a spatial conflict over who can access functioning institutions and who is trapped by their failure.

\subsection{Beyond Inequality: The Crisis of Capacity and Paralysis}
The prevailing academic explanation relies heavily on the distributional perspective, epitomized by the Stolper-Samuelson theorem (1941). This narrative posits that globalization suppressed the wages of low-skilled workers in developed nations, leading to a populist backlash (Autor, Dorn, \& Hanson, 2013, 2016).

While valid, this explanation is critically insufficient because it fundamentally misdiagnoses the nature of the current crisis. It explains the economic \textit{grievance} but fails to explain the \textit{systemic paralysis} of state capacity and the pervasive sense of institutional \textit{chaos}. The distinctive feature of the current era is not merely anxiety over income, but profound institutional dysfunction—characterized by administrative gridlock, logistical breakdowns, and polarized governance.

If the problem were solely income inequality, standard political economy models would predict mitigation through linear transfer payments (e.g., Trade Adjustment Assistance). Yet, these mechanisms have proven inadequate, and crucially, the capacity to implement them effectively has eroded. The empirical reality we observe—such as the 2021-2022 supply chain crisis or the exponential backlog in immigration courts—points to a deeper structural failure.

This failure stems from a critical blind spot in traditional models, which often assume infinite or homogeneous institutional capacity. By treating the costs of globalization merely as a matter of ``how to slice the pie,'' these models overlook the binding constraint: the stagnation of the state's institutional capacity to manage the flows of trade, capital, and migration.

We argue for a fundamental theoretical shift: moving beyond the traditional focus on the \textit{distribution of income} to analyze the \textit{distribution of institutional capacity and congestion costs}. The crisis of globalization is fundamentally a spatial conflict over access to functioning institutions in an era of systemic overload.
\subsection{A New Framework: Optimization Cutoffs and Heterogeneous Capacity}
We propose a novel framework that reimagines the national economy as a ``Congestible Club Good'' \citep{buchanan1965}. Globalization flows ($M$)—trade, migration, capital—are not costless; they must be processed by domestic Institutional Capacity ($K$). When $M$ exceeds $K$, the system experiences congestion externalities—social friction, administrative paralysis, and chaos—that rise exponentially.

Crucially, we argue that capacity is neither stagnant by accident nor homogeneous in its distribution. We introduce two critical concepts to explain the US predicament:

\subsubsection*{The Mechanism of Stagnation: The ``Optimization Cutoff''}
We must first explain \textit{why} the hegemon allowed its capacity to stagnate. We introduce the concept of the ``Optimization Cutoff.'' Consider the analogy of an aging urban center, like Beijing's inner city. While inefficient infrastructure (e.g., dense ``Hutongs'' or urban villages) persists within the core, this is not necessarily irrational. Renovating the old core is exceedingly expensive and politically complex.

When the option to develop the ``suburbs'' is available at a lower marginal cost, capital naturally flows to the path of least resistance. Globalization acted as the ``suburb'' for the US economy. For decades, investing in global supply chains and leveraging emerging markets' infrastructure offered higher returns and lower costs than the arduous task of upgrading aging US infrastructure (the ``Rust Belt'') or reforming complex domestic institutions.

Consequently, the path of internal optimization was prematurely \textit{cut off}. Capital (both public and private) rationally bypassed domestic structural problems. This led to a path dependency characterized by chronic underinvestment in the domestic core, even as the overall global system expanded.

\subsubsection*{The Structure of Polarization: Heterogeneous Capacity}
This long-term neglect did not affect all citizens equally. We disaggregate Institutional Capacity ($K$) into two distinct components:
\begin{itemize}
    \item \textbf{Public Capacity ($K_{Public}$):} The universally accessible infrastructure and governance systems (ports, public courts, social safety nets). Due to the Optimization Cutoff, this capacity has stagnated.
    \item \textbf{Private Capacity ($K_{Private}$):} The exclusive resources accessible to elites (private education, customized logistics, specialized legal services). This capacity remains efficient and has expanded.
\end{itemize}

The polarization over globalization stems directly from this capacity differential. When global flows ($M$) surged, they collided with a fragile and stagnant $K_{Public}$, creating two distinct political constituencies:

\begin{itemize}
    \item \textbf{The Insulated Elites (High $K_{Private}$):} This group benefits significantly from globalization flows ($M$) and can utilize $K_{Private}$ to bypass the congestion in the public system. For them, the optimization space remains large ($MB > MC$). They rationally support openness and often oppose costly investments in $K_{Public}$.
    \item \textbf{The Congested Incumbents (Dependent on $K_{Public}$):} This group relies on the stagnant public infrastructure. They directly bear the brunt of the congestion—experiencing globalization as overwhelmed local services, infrastructural failure, and social chaos ($M > K_{Public}$). Their optimization space has vanished ($MC > MB$). Their rational response is to demand protectionism.
\end{itemize}

The ``Great Reversal'' is thus driven by the political activation of the Congested Incumbents, whose localized experience of overload has finally dominated the national political agenda.

\subsection{Methodology and Key Findings}
Methodologically, this study combines a theoretical simulation with empirical textual analysis.

First, our numerical simulation models the impact of the ``Optimization Cutoff'' and the resulting capacity heterogeneity. It demonstrates that due to the failure of $K_{Public}$, a large segment of the US population is trapped in a ``Dis-economy Zone.'' Furthermore, the model reveals a ``Capacity Trap'': the polarization generated by uneven congestion paralyzes the political consensus required to restore $K_{Public}$ (as elites resist funding a system they have exited), locking the system into a suboptimal equilibrium of protectionism.

Second, utilizing the LexisNexis database, we construct an ``Institutional Congestion Index'' based on US mainstream media and congressional documents from 2000 to 2024. The empirical results reveal a startling exponential surge in disorder-related keywords (e.g., ``chaos,'' ``overwhelmed'') associated with globalization flows—rising from 272 hits in 2000 to 1,333 in 2024. We interpret this as the political manifestation of $K_{Public}$ overload.

Third, we triangulate this perception of chaos with hard evidence of physical overload: the spike in the Global Supply Chain Pressure Index (GSCPI) and the explosion of the US Immigration Court backlog to 3.7 million cases. This provides robust evidence that the US public institutional capacity has breached its critical threshold.
\subsection{Roadmap of the Paper}
The remainder of this paper is organized as follows. Section 2 situates our framework within the broader IPE literature, arguing that capacity is the critical variable mediating distributional and geopolitical pressures. Section 3 establishes the theoretical model of heterogeneous institutional congestion and the dynamics of the Optimization Cutoff. Section 4 presents the simulation calibration and analyzes the political economy of the ``Capacity Trap.'' Section 5 provides the empirical evidence and triangulation. Section 6 discusses the policy implications for the US and China, and concludes.

\section{Theoretical Dialogue: Beyond Motives, \texorpdfstring{\\}{ }Towards Mechanisms of Collapse}
To situate our ``Fractured Metropolis'' framework within the broader academic debate, we must engage with the dominant narratives explaining the ``Great Reversal'' of US globalization policy. The existing literature provides crucial insights into the \textit{motives} for retrenchment—ranging from economic inequality to geopolitical rivalry.

However, we argue that these approaches collectively fail to account for the specific \textit{mechanisms} of systemic dysfunction and the profound internal polarization that characterize the current crisis. The empirical reality we observe (detailed in Section 5) is not merely a policy shift, but a non-linear surge in institutional ``chaos'' and a spatial fracturing of the political landscape. Existing theories, often assuming stable and homogeneous institutional capacity, struggle to explain these dynamics.

We synthesize and critique three dominant IPE narratives—Distributional, Geopolitical, and Institutional—to demonstrate that the historical process of capacity stagnation (the Optimization Cutoff) and the resulting heterogeneity of capacity ($K$) are the critical variables mediating macro-level pressures and the micro-dynamics of systemic overload.

\subsection{The Distributional Narrative and the Crisis of Capacity Inequality}
The most prevalent explanation for the backlash against globalization is grounded in the political economy of distribution. Building on the Stolper-Samuelson theorem (1941), this view posits that trade integration suppresses the real wages of low-skilled workers in capital-abundant economies. Seminal work on the ``China Shock'' \citep{autor2013, autor2016} provides robust evidence linking import competition to localized economic distress.

While the distributional narrative powerfully explains the \textit{grievance} of certain segments of the population (our ``Congested Incumbents''), it struggles to explain the \textit{paralysis} of state capacity and the depth of the political polarization.

\subsubsection*{The Missing Link: From Income to Capacity}
The standard narrative assumes the state \textit{could} mitigate the impact through redistribution, consistent with the framework of ``Embedded Liberalism'' \citep{ruggie1982}. The critical failure lies in the inability of the US political system to implement these solutions effectively.

Our framework argues this failure stems from a deeper crisis of \textit{capacity inequality}. We move beyond income distribution to the distribution of access to functioning institutions. The distributional narrative implicitly assumes that all actors rely on the same public capacity ($K_{Public}$). However, the ability of elites to ``exit'' into private capacity ($K_{Private}$) fundamentally alters the political equilibrium.

Polarization deepens not just because some lose income, but because the ``Congested Incumbents'' are trapped in deteriorating public systems (infrastructure, social services) while the ``Insulated Elites'' bypass them. This heterogeneity explains the political paralysis: elites (high $K_{Private}$) have reduced incentives to fund the modernization of $K_{Public}$.

\subsubsection*{The Historical Blind Spot: The Optimization Cutoff}
Furthermore, the distributional literature fails to adequately explain \textit{why} the state's buffering capacity eroded. We argue this erosion was endogenous to globalization itself. The ``Optimization Cutoff'' mechanism demonstrates that globalization incentivized the neglect of $K_{Public}$.

As capital rationally prioritized global expansion (the ``suburbs'') over costly domestic upgrades (the ``inner city''), the institutional machinery required to manage distributional conflicts was systematically underfunded. The crisis, therefore, is not just about how to slice the pie, but about the shrinking and fractured capacity of the institutional machinery that manages the system.
\subsection{The Geopolitical Narrative and the Fractured State}
A second strand of literature, rooted in Realism and Hegemonic Stability Theory (HST) \citep{gilpin1981, kindleberger1973}, views the US retreat as a rational grand strategy. The US is withdrawing from the open order because it now disproportionately benefits its primary rival, China. Protectionism is interpreted as a deliberate attempt to ``weaponize interdependence'' \citep{farrell2019}.

While this narrative captures the external logic of great power competition, it suffers from a critical flaw: it treats the state as a unitary, rational actor. This overlooks the profound internal dysfunction and polarization characterizing the US response. The chaotic implementation of tariffs, the reactive crisis management, and the logistical breakdowns suggest a system under duress rather than one executing a calculated strategy.

\subsubsection*{Internalizing Hegemonic Costs}
Our framework challenges the traditional understanding of hegemonic costs within HST. Traditionally, these costs are viewed as external—providing global public goods. We emphasize \textit{Internal Processing Costs}: the domestic congestion incurred by serving as the central node of the global economy.

The US is not retreating solely because it fears a rival, but because its own public capacity ($K_{Public}$) is saturated. The Optimization Cutoff explains this vulnerability: the long-term prioritization of global reach over core resilience created a brittle domestic system. As $K_{Public}$ declines, the domestic political cost of maintaining global openness rises exponentially. The retreat looks less like a grand strategy and more like emergency triage to manage internal chaos.

\subsubsection*{The Spatial Mismatch}
The geopolitical friction is driven by a fundamental spatial mismatch rooted in asymmetric capacity cycles. The US (the saturated ``inner city'') needs to reduce flow ($M$) to manage internal congestion. China (the expansive ``suburb'') possesses surplus capacity and needs to increase flow. This dynamic, rather than mere strategic competition, is a fundamental driver of the current conflict.

\subsection{The Institutional Narrative: Dynamizing and Localizing the Trilemma}
The third narrative focuses on the fundamental incompatibility of deep integration with national sovereignty. Dani Rodrik’s (2011) ``Globalization Trilemma'' famously argues that nations cannot simultaneously pursue democracy, national sovereignty, and hyper-globalization. As integration deepens, it clashes with domestic social contracts \citep{rodrik2018}.

Our paper accepts the Trilemma as a fundamental constraint but seeks to explain the \textit{timing} of the collapse and \textit{who} experiences the tension most acutely. Rodrik’s framework is structurally static; it defines the trade-offs but not the tipping points or the internal distribution of the burden.

\subsubsection*{Capacity as the Dynamic Variable}
We introduce Institutional Capacity ($K$) as the dynamic variable that conditions the Trilemma. When $K$ is high, a state possesses the ``institutional buffer'' to manage the friction between globalization and democracy. The Optimization Cutoff explains the historical dynamic: the long-term erosion of the US institutional buffer eventually triggered the crisis.

\subsubsection*{The Heterogeneity of the Trilemma}
Crucially, the heterogeneity of capacity means the Trilemma does not bind uniformly across society. Elites possessing high $K_{Private}$ can effectively navigate the Trilemma; they enjoy the benefits of hyper-globalization while remaining insulated from its social costs.

The Trilemma binds most tightly on the ``Congested Incumbents'' who rely on $K_{Public}$. They experience the clash acutely, as globalization undermines the public services and social contracts they depend upon. The political explosion (populism) occurs when this group mobilizes. Our framework suggests that as capacity is breached, the political and social costs of managing the Trilemma rise exponentially, explaining the suddenness of the ``Great Reversal.''

\subsection{Synthesis: The Political Economy of Spatial Congestion}
By synthesizing these perspectives, we propose a unified framework. We accept the distributional grievances (Narrative 1) and the geopolitical frictions (Narrative 2), but we demonstrate that both are amplified by, and often symptomatic of, a deeper underlying cause: the physical and governance limits of a \textit{Fractured Metropolis}.

The ``Great Reversal'' is a systemic response to uneven Institutional Congestion. This perspective offers a novel contribution to IPE by fundamentally redefining how we analyze globalization. It shifts the focus from globalization as a \textit{policy choice}—evaluated primarily through lenses of fairness or strategic advantage—to globalization as a \textit{flow} that must be physically and institutionally processed. Furthermore, it reveals that the capacity to process this flow is not only finite but deeply unequal in its distribution, creating the spatial political conflicts that define the current era.
\section{Theoretical Framework: A Spatial Model of Institutional Congestion}
This section develops a theoretical framework to understand how institutional capacity shapes a nation's optimal level of globalization and how the historical evolution and distribution of that capacity drive the current political crisis. We begin with the basic setup of the economy as a congestible facility, introduce the historical mechanism that explains capacity stagnation—the Optimization Cutoff—and finally model how this stagnation leads to a ``Fractured Metropolis'' characterized by heterogeneous capacity and political polarization.

\subsection{The Economy as a Congestible Facility: Basic Setup}
Drawing upon the theory of club goods \citep{buchanan1965} and classic congestion theory \citep{vickrey1969}, we conceptualize the national economy not as a boundless open system, but as a ``congestible facility'' with finite capacity. We define two primary variables:

\begin{itemize}
    \item \textbf{Institutional Capacity ($K$):} The aggregate capacity of a country to process external flows. This includes physical infrastructure, administrative efficiency, and social governance systems.
    \item \textbf{Globalization Intensity ($M$):} The volume of external flows (trade, capital mobility, migration) impacting the domestic system.
\end{itemize}

We argue that maintaining order in an open economy requires consuming $K$. The objective is to maximize Net National Welfare ($W$), defined as the economic benefits from globalization ($B(M)$) minus the social congestion costs ($C(M)$).

\subsubsection*{The Benefit Function $B(M)$}
We assume globalization brings economic gains. Following standard models \citep[e.g.,][]{arkolakis2012}, $B(M)$ is concave, exhibiting diminishing marginal returns. We also incorporate a \textbf{``Catch-up Parameter'' ($\delta$)}. For frontier economies (US), $\delta \approx 0$; for developing economies (China), $\delta > 0$. This parameter is essential for understanding why latecomers have a higher tolerance for intensive globalization flows.

\subsubsection*{The Congestion Cost Function $C(M)$}
The core of our model lies in how we conceptualize the costs. We posit that the cost of maintaining social order is a function of the \textbf{``Congestion Ratio'' ($M/K$)}:

\begin{equation}
    C(M) = \gamma \left( \frac{M}{K} \right)^\phi
\end{equation}

Where $\gamma$ is the sensitivity to disorder, and $\phi$ is the Congestion Elasticity. The crucial assumption is that $\textbf{$\phi > 1$}$. This implies the cost function is strictly convex: as globalization intensity increases linearly, the social and institutional costs rise \textit{exponentially}.

\subsubsection*{The Intuition of $\phi > 1$: Political Externalities and Governance Decay}
Why do costs rise exponentially? We argue that institutional congestion generates severe \textit{Political Externalities} that scale non-linearly. When a system exceeds its capacity threshold, it does not merely slow down (linear cost); it gridlocks and begins to fail (exponential cost).

We draw on the characterization of infrastructure and governance as ``Chaotic Sociotechnical Systems'' \citep{el-diraby2013}. However, the non-linearity extends beyond mere logistical bottlenecks (e.g., port saturation leading to warehouse shortages, which further block ports). Crucially, the overload of institutional capacity triggers \textbf{self-reinforcing feedback loops of governance decay}.

When public services (like courts or administrative agencies) are overwhelmed, it degrades social trust and violates the implicit social contract. This failure fuels political polarization as different groups clash over increasingly scarce institutional resources. This polarization, in turn, paralyzes the administrative capacity needed to resolve the initial congestion. This dynamic—where congestion breeds polarization, which further degrades governance capacity—explains why the political costs of disorder scale much faster than the flow itself.
\subsection{The Historical Dynamics of Capacity: The Optimization Cutoff}
Before analyzing the current equilibrium, we must explain \textit{why} the institutional capacity of the hegemon ($K_{US}$) stagnated. We argue this stagnation was not an oversight but a rational, path-dependent outcome of globalization itself. We introduce the mechanism of the \textbf{``Optimization Cutoff.''}

During the peak era of globalization (c. 1980-2008), the US economy faced a fundamental choice regarding resource allocation, akin to an expanding metropolis:

\begin{enumerate}
    \item \textbf{Internal Optimization (The ``Inner City'' Path):} Upgrading aging domestic infrastructure, reforming complex social institutions, and strengthening safety nets.
    \item \textbf{External Expansion (The ``Suburb'' Path):} Leveraging the infrastructure and labor of emerging markets and expanding global supply chains.
\end{enumerate}

\textbf{The Cutoff Mechanism:} The critical insight is that the Internal path faced significantly higher marginal costs ($MC_{Internal}$) than the External path ($MC_{External}$). Crucially, these internal costs were not just economic; they were also \textit{political}. Renovating the domestic core required navigating dense regulatory complexity, overcoming entrenched interests, and mobilizing sustained political consensus—all high-friction activities.

In contrast, global expansion offered lower economic costs and, importantly, \textbf{lower political resistance}. It allowed capital (both private investment and public policy focus) to rationally bypass these costly domestic structural problems. Consequently, the path of internal optimization was prematurely \textit{cut off} long before reaching its potential maximum.

\textbf{Path Dependency and Stagnation:} Globalization acted as a critical juncture that locked the US into a \textit{Path Dependency} characterized by chronic underinvestment in the domestic core. This long-term ``systemic bypass'' explains the fragility of $K_{Public}$. The persistence of the ``Rust Belt'' or aging infrastructure (analogous to ``Hutongs'' or urban villages persisting within a modern city) is the physical manifestation of this cutoff.

Globalization, therefore, endogenously incentivized the neglect of the very domestic capacity required to sustain it in the long run. When the massive flows generated by this expansion eventually collided with the neglected domestic core, the system overloaded.
\subsection{The Fractured Metropolis: Heterogeneous Capacity and Polarization}
The Optimization Cutoff did not affect all segments of society equally. The long-term neglect of domestic capacity led to a critical divergence in how citizens access the benefits of globalization and experience its costs. To capture this, we disaggregate Institutional Capacity ($K$) into two distinct components:

\begin{itemize}
    \item \textbf{$K_{Public}$ (Public Capacity):} Universally accessible infrastructure (roads, ports) and governance systems (public courts, social safety nets). This is the capacity primarily affected by the Optimization Cutoff, leading to stagnation and fragility.
    \item \textbf{$K_{Private}$ (Private/Club Capacity):} Exclusive resources accessible to elites, allowing them to bypass public bottlenecks (e.g., private education, customized logistics, specialized legal and financial services). This capacity remained efficient and expanded.
\end{itemize}

The divergence between $K_{Public}$ and $K_{Private}$ creates the structure of the ``Fractured Metropolis'' and defines the two key political constituencies:

\subsubsection*{The Insulated Elites (Access to $K_{Private}$)}
This group derives significant economic benefits from globalization flows ($M$). Crucially, they possess sufficient $K_{Private}$ to insulate themselves from the congestion in the public system.
\begin{itemize}
    \item \textbf{Experience:} $MB > MC$. They navigate the system efficiently. Their optimization space remains large.
    \item \textbf{Political Preference:} Rationally support openness. Furthermore, because they have effectively ``exited'' the public system, they often oppose costly investments required to upgrade $K_{Public}$ (e.g., resisting tax increases for infrastructure they do not primarily use).
\end{itemize}

\subsubsection*{The Congested Incumbents (Dependent on $K_{Public}$)}
This group relies heavily on the stagnant and fragile public infrastructure. When global flows ($M$) surge, they directly bear the brunt of the congestion.
\begin{itemize}
    \item \textbf{Experience:} $MC > MB$. They experience globalization as chaos, overwhelmed local services, and infrastructural failure. Their Congestion Ratio ($M/K_{Public}$) is critically high. Their optimization space is exhausted.
    \item \textbf{Political Preference:} Rationally demand protectionism and ``load shedding'' (reducing $M$).
\end{itemize}

\textbf{The Dynamics of Polarization:} The political battle over globalization is fundamentally a conflict between these two realities. Polarization is endogenous to the heterogeneous distribution of congestion. The crisis occurs when the exponential costs ($\phi > 1$) hitting the Congested Incumbents become so severe that their political mobilization overwhelms the traditional pro-globalization consensus.

\subsection{Synthesis: Equilibrium and Key Propositions}
The aggregate national optimal level of globalization ($M^*$) is determined by maximizing the Net National Welfare function $W(M)$, balancing the marginal benefits and the marginal congestion costs. (The full derivation of the equilibrium and the comparative statics are provided in Appendix A).

The equilibrium analysis reveals how the mechanisms described above lead to the divergent paths of the US and China, and why the US is locked into a suboptimal outcome.

\subsubsection*{Proposition 1: Asymmetric Paths (The Capacity Constraint)}
The optimal level of globalization ($M^*$) is strictly increasing in Institutional Capacity ($K$) and the Catch-up Parameter ($\delta$). This explains the structural divergence in the global economy:
\begin{itemize}
    \item \textbf{China (Expanding K, High $\delta$):} Benefiting from massive infrastructure investment (avoiding the Optimization Cutoff) and the catch-up effect, China resides in the ``Climbing Phase.'' Its $M^*$ is high, and the current $M < M^*$. Expansion is rational.
    \item \textbf{US (Stagnant K, Low $\delta$):} Facing capacity stagnation (due to the Optimization Cutoff) and being at the technological frontier, the US has a lower $M^*$. The current $M > M^*$, pushing the economy into the ``Dis-economy Zone.''
\end{itemize}

\subsubsection*{Proposition 2: The Capacity Trap (The Political Economy of Stagnation)}
The model reveals a critical political failure driven by heterogeneous capacity. While increasing $K$ (specifically $K_{Public}$) is the \textbf{First-Best} economic solution for the US, the system defaults to the \textbf{Second-Best} solution: protectionism (reducing $M$).

This ``Capacity Trap'' emerges because the polarization between the Insulated Elites and the Congested Incumbents paralyzes the political consensus required for large-scale public investment. Elites resist funding $K_{Public}$, while Incumbents demand immediate relief from congestion. In this environment of paralyzed governance, reducing $M$ (e.g., via tariffs or border restrictions) becomes the path of least political resistance, locking the US into an equilibrium of stagnation and retreat.

\section{Numerical Simulation and Calibration}

\subsection{Parameter Calibration: The Empirical Shadow of the Optimization Cutoff}

To numerically solve the welfare optimization problem derived in Section 3 (and Appendix A), we calibrate the model parameters by synthesizing macro-level stylized facts with authoritative empirical data. Unlike standard models, we explicitly model Institutional Capacity ($K$) as a dynamic constraint shaped by historical investment choices. Our calibration of the US-China asymmetry relies on a Multi-dimensional Assessment of capacity dynamics, drawing evidence from the OECD, the US Bureau of Economic Analysis (BEA), and the World Bank (WGI). The baseline parameter values are summarized in Table 1, and the supporting stylized facts justifying the capacity gap are visualized in Figure 1.

\begin{figure}[H]
    \centering
    \includegraphics[width=\textwidth]{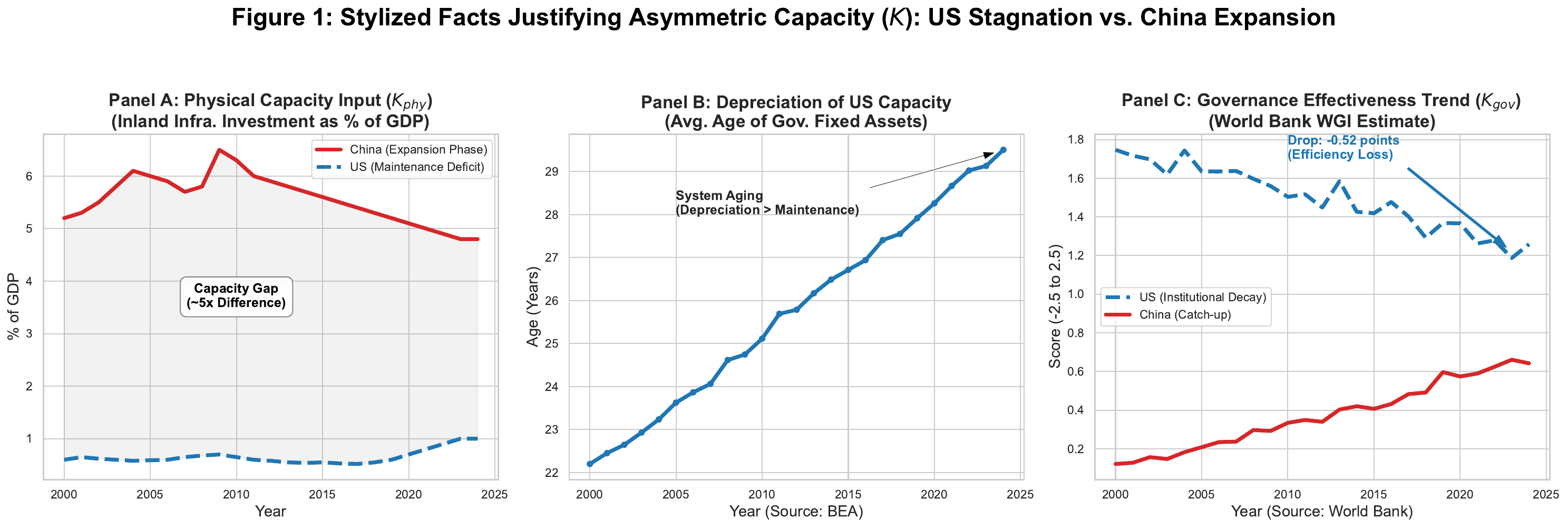}
    \caption{Stylized Facts Justifying Asymmetric Capacity ($K$): US Stagnation vs. China Expansion}
    \label{fig:stylized_facts}
\end{figure}

\subsubsection{Institutional Capacity (\texorpdfstring{$K$}{K}): The Legacy of the Optimization Cutoff}

The most critical parameter is $K$, representing the \textbf{Effective Throughput Capacity} of an economy. In the context of our model, the binding constraint on globalization is primarily the capacity of the public infrastructure and governance systems ($K_{Public}$). We calibrate $K_{CN} = 8.0$ (High Capacity/Expansion) and $K_{US} = 5.0$ (Low Capacity/Stagnation). This calibration is grounded in a striking divergence observed across both physical ($K_{phy}$) and governance ($K_{gov}$) dimensions (Figure 1). We interpret this divergence as the empirical manifestation of the \textbf{``Optimization Cutoff''} mechanism introduced in Section 3.2.

\paragraph{The Physical Dimension ($K_{phy}$): The ``Investment Scissors'' and Systemic Bypass}
Capacity requires continuous investment. Figure 1, Panel A, reveals a persistent ``Investment Scissors.'' Since 2000, China’s inland infrastructure investment averaged approximately 5.8\% of GDP. In contrast, the US stagnated between 0.6\% and 1.0\% (OECD, 2024).

\textbf{Interpretation (The Optimization Cutoff):} This massive gap is not accidental. It reflects the historical period where the marginal cost of upgrading the US domestic core ($MC_{Internal}$) exceeded the cost of global expansion ($MC_{External}$). Capital rationally bypassed domestic investment in favor of global supply chains. This systematic neglect led to the current fragility of $K_{Public}$.

\paragraph{The Consequence: Net Depreciation and Aging Stock}
The long-term effect of the Optimization Cutoff is the physical degradation of the domestic core. As shown in Figure \ref{fig:stylized_facts}, Panel B, the Average Age of Government Fixed Assets in the US has risen monotonically from 22.1 years in 2000 to 29.5 years in 2024 (BEA, 2024). This aging stock physically constrains the system's efficiency.

\paragraph{The Governance Dimension ($K_{gov}$): Institutional Decay}
As illustrated in Figure 1, Panel C, the US score for ``Government Effectiveness'' (WGI) has declined significantly from 1.74 (2004) to 1.22 (2023) (World Bank, 2024).

\textbf{Interpretation:} The \textbf{negative trajectory} signals ``Institutional Decay.'' This persistent deterioration violates social expectations, fueling the polarization that paralyzes the decision-making process required to maintain K (The ``Capacity Trap,'' detailed in Section 4.5).

\paragraph{Calibration Logic:}
The calibration $K_{US} < K_{CN}$ synthesizes these factors. The massive deficit in physical investment (the legacy of the Optimization Cutoff) places a hard constraint on US public capacity. We conservatively set the effective capacity ratio at 1.6:1 ($K_{CN}=8$ vs. $K_{US}=5$).
\subsubsection{Other Parameters (\texorpdfstring{$\phi, \alpha, \theta, \delta, \gamma$}{phi, alpha, theta, delta, gamma})}

\paragraph{Congestion Elasticity ($\phi$): The Non-linear Threshold}
We set the congestion elasticity parameter $\phi = 2.5$. This implies that the cost function $C(M) = \gamma (M/K)^\phi$ is strictly convex.

\textbf{Theoretical Basis:} This draws on El-Diraby’s (2013) characterization of civil infrastructure as a \textbf{``Chaotic Sociotechnical System,''} where stress beyond a tipping point leads to non-linear feedback loops rather than linear delays.

\textbf{Empirical Consistency:} This parameter choice is consistent with the ``hockey stick'' trajectory of our empirical Institutional Congestion Index, which shows disorder keywords surging exponentially after 2016.

\paragraph{Benefit Parameters ($\alpha, \theta, \delta$) and Social Sensitivity ($\gamma$)}
\begin{itemize}
    \item \textbf{Benefit Elasticity ($\theta = 0.6$):} Following Arkolakis et al. (2012), we assume diminishing marginal returns to trade openness.
    \item \textbf{Baseline Technology ($\alpha$):} We set $\alpha_{US} = 2.0$ and $\alpha_{CN} = 1.0$. This reflects the productivity gap documented in the Penn World Table (PWT 10.0; Feenstra et al., 2015).
    \item \textbf{Catch-up Parameter ($\delta$):} We set $\delta_{US} = 0$ and $\delta_{CN} = 0.5$. This captures the technology spillover effect specific to latecomers.
    \item \textbf{Social Sensitivity ($\gamma$):} We normalize $\gamma = 1.0$ for both countries. While the discussion on the ``Expectations Gap'' might suggest the US currently has a higher sensitivity to capacity decline ($\gamma_{US} > \gamma_{CN}$), we maintain standardization in the baseline model. This conservative approach allows the model's divergent outcomes to be driven purely by structural differences in capacity ($K$) and catch-up potential ($\delta$), rather than assumptions about societal preferences.
\end{itemize}

\begin{table}[h]
    \centering
    \caption{Baseline Parameter Calibration and Justification}
    \label{tab:calibration}
    \begin{tabular}{|p{3cm}|c|c|c|p{6cm}|}
        \hline
        \textbf{Parameter} & \textbf{Symbol} & \textbf{Value (US)} & \textbf{Value (CN)} & \textbf{Source / Justification} \\
        \hline
        \multicolumn{5}{|c|}{\textit{Benefit Parameters}} \\
        \hline
        Baseline Technology & $\alpha$ & 2.0 & 1.0 & Reflects the productivity gap based on Total Factor Productivity (TFP) levels (PWT 10.0 stylized facts). \\
        \hline
        Technology Catch-up & $\delta$ & 0.0 & 0.5 & Captures technology spillovers for latecomers. US (Frontier) has $\delta=0$. Calibrated based on TFP \textbf{growth} differentials. \\
        \hline
        Benefit Elasticity & $\theta$ & 0.6 & 0.6 & Standard assumption of diminishing returns to trade openness (Arkolakis et al., 2012). \\
        \hline
        \multicolumn{5}{|c|}{\textit{Cost Parameters}} \\
        \hline
        Institutional Capacity & $K$ & 5.0 & 8.0 & \textbf{Multi-dimensional Assessment.} Based on divergent trajectories in physical investment (OECD), asset depreciation (BEA), and governance trends (WGI). \\
        \hline
        Congestion Elasticity & $\phi$ & 2.5 & 2.5 & ``Chaotic Sociotechnical System'' theory (El-Diraby, 2013); consistent with exponential empirical trend. \\
        \hline
        Social Sensitivity & $\gamma$ & 1.0 & 1.0 & Standardization/Normalization. Isolates the effects of structural differences in $K$ and $\delta$. \\
        \hline
    \end{tabular}
\end{table}

\subsection{Simulation Results: Asymmetric Paths under Globalization}

\subsubsection*{4.2.1 Calibration of Globalization Intensity ($M$): The Convergence Hypothesis}
The variable $M$ represents a normalized index of globalization intensity (scaled 0 to 10). Based on indices measuring ``Globalization Depth'' \citep{dhl2024}, the globalization intensities of the US and China have significantly converged. We calibrate the current globalization intensity for both countries at $M_{US} \approx M_{CN} \approx 6.0$.

\textbf{The Significance of Convergence:} By holding the external shock constant ($M=6.0$), we isolate the impact of domestic capacity ($K$) and the Catch-up Effect ($\delta$). This highlights the central mechanism of the ``Congestion Ratio'' ($M/K$):
\begin{itemize}
    \item \textbf{US Overload:} With $K_{US}=5.0$ (stagnated $K_{Public}$), the ratio is $6.0/5.0 = 1.20$. The load exceeds capacity.
    \item \textbf{China Buffer:} With $K_{CN}=8.0$, the ratio is $6.0/8.0 = 0.75$. The load is within capacity limits.
\end{itemize}

\subsubsection*{4.2.2 Aggregate Simulation Results: The Great Divergence}
\begin{figure}[H]
    \centering
    \includegraphics[width=\textwidth]{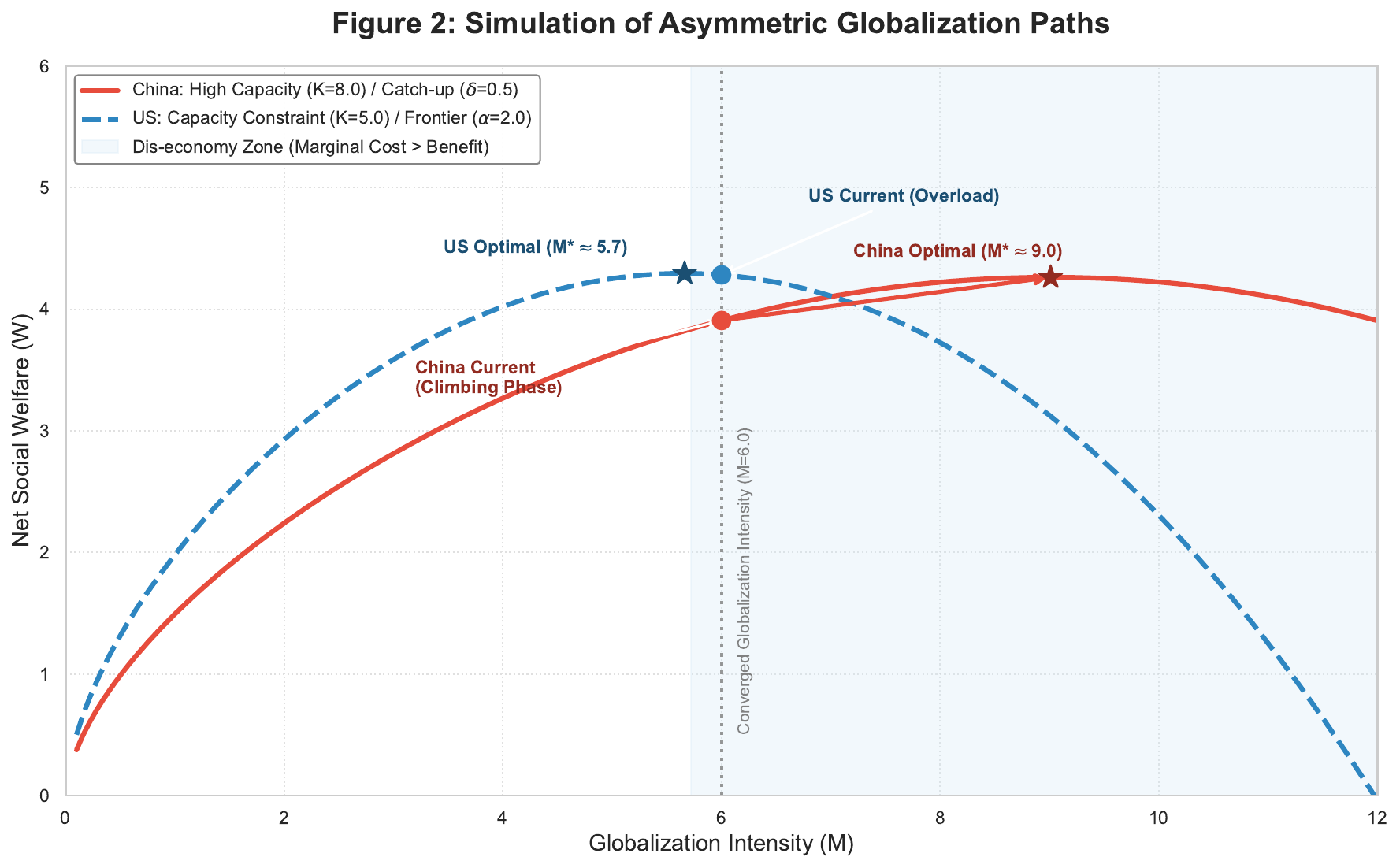} 
    \caption{Simulation of Asymmetric Globalization Paths. \textit{Blue Dashed Line}: US (Capacity Constraint); \textit{Red Solid Line}: China (High Capacity). The US is currently in the ``Dis-economy Zone'' ($M > M^*$).}
    \label{fig:simulation_results}
\end{figure}

Figure \ref{fig:simulation_results} illustrates the simulated net welfare curves. The results demonstrate how identical globalization intensity leads to diametrically opposed outcomes.

\begin{itemize}
    \item \textbf{The US Scenario (The ``Overload'' Trap):} Due to lower capacity ($K=5.0$), the welfare curve peaks early at $M^*_{US} \approx 5.7$. Since actual intensity ($M=6.0$) is to the \textit{right} of the peak ($MC > MB$), the rational aggregate strategy is protectionism.
    \item \textbf{The China Scenario (The ``Climbing'' Phase):} Thanks to high capacity ($K=8.0$) and catch-up ($\delta=0.5$), the curve peaks later at $M^*_{CN} \approx 9.0$. At $M=6.0$, China resides on the \textit{left} side, incentivizing continued openness.
\end{itemize}

\subsection{Discussion: The Structural Mismatch and the Legacy of the Optimization Cutoff}
The simulation results highlight a fundamental theoretical insight: Globalization is not a scalable, one-size-fits-all remedy; it is a conditional variable constrained by domestic architecture.

\subsubsection*{The Endogeneity of Optimal Openness}
Standard trade theories often treat institutional capacity ($K$) as exogenous. Our model refutes this. We demonstrate that $M^*$ is strictly endogenous to $K$. The stagnation of $K$ in the US is the direct legacy of the \textbf{``Optimization Cutoff.''} By historically prioritizing global expansion over domestic upgrades, the US system allowed $K_{Public}$ to erode. When this neglected capacity collides with high flows, the ``Congestion Cost'' dominates. The ``Great Reversal'' is thus a predictable systemic response to the violation of the condition $M \le K_{Public}$.

\subsubsection*{The Asymmetry of Strategic Time Zones}
The divergence reflects a deep ``temporal misalignment.''
\begin{itemize}
    \item \textbf{The US ``Capacity Deficit'' (The Saturated Inner City):} The US resides in the ``Post-Maturity'' phase. With fragile infrastructure ($K \approx 5.0$) and high intensity ($M \approx 6.0$), every unit of external shock is amplified into a political crisis.
    \item \textbf{The China ``Capacity Surplus'' (The Expanding Suburb):} China operates in the ``Accumulation'' phase. Its massive infrastructure ($K \approx 8.0$) creates a buffer to absorb flows profitably.
\end{itemize}

\subsubsection*{From Misalignment to Conflict: The Spatial Clash}
Critically, this asymmetry creates a structural driver for geopolitical conflict. The US (saturated inner city) is rationally motivated to implement \textbf{``Zoning''} (tariffs, sanctions) to relieve internal congestion. China (expansive suburb) is rationally motivated to maximize \textbf{``Connectivity''} to utilize surplus capacity. This is not merely a competition for market share; it is a clash of systemic lifecycles.
\subsection{Robustness: Sensitivity Analysis of Key Parameters}
\begin{figure}[H]
    \centering
    \includegraphics[width=\textwidth]{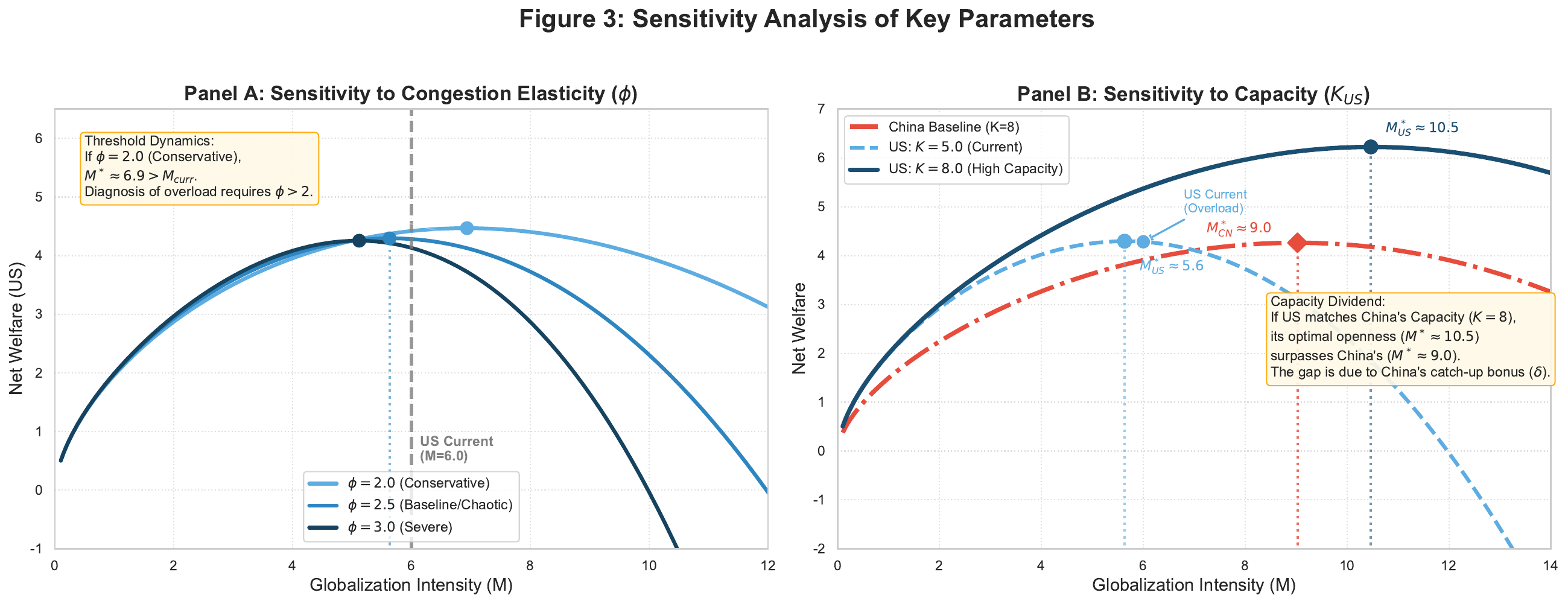} 
    \caption{Sensitivity Analysis of Key Parameters. \textit{Panel A}: Sensitivity to Congestion Elasticity ($\phi$); \textit{Panel B}: Sensitivity to Capacity ($K_{US}$).}
    \label{fig:sensitivity}
\end{figure}

Our baseline simulation relies on calibrated parameters. To ensure that our core diagnosis—the ``Institutional Overload''—is robust, we perform a systematic sensitivity analysis on the two most critical variables: the congestion elasticity ($\phi$) and the institutional capacity ($K$). Figure \ref{fig:sensitivity} presents the results.

\subsubsection*{4.4.1 Sensitivity to Congestion Elasticity ($\phi$): The ``Razor's Edge''}
In the baseline model, we set $\phi = 2.5$ to capture the non-linear, ``chaotic'' nature of socio-technical systems \citep{el-diraby2013}. Panel A of Figure \ref{fig:sensitivity} examines the system's behavior across a range of congestion regimes.

\begin{itemize}
    \item \textbf{The Conservative Scenario ($\phi = 2.0$):} If we assume a standard quadratic cost function, the theoretical optimal peak for the US is $M^* \approx 6.9$. Since this is higher than the current intensity ($M \approx 6.0$), the US would still be in a ``safe zone.''
    \item \textbf{The Chaotic Reality ($\phi \ge 2.5$):} However, once we account for the feedback loops characteristic of complex institutional systems, the optimal threshold shifts left. At our baseline ($\phi=2.5$), the peak drops to $M^* \approx 5.7$.
\end{itemize}

\textbf{Interpretation:} Since the current intensity ($M \approx 6.0$) exceeds this threshold ($6.0 > 5.7$), the diagnosis of ``System Overload'' is confirmed. This underscores that the crisis is driven by the specific \textit{non-linear nature} of institutional failure—chaos scales faster than flow.

\subsubsection*{4.4.2 Sensitivity to Capacity ($K$): The Opportunity Cost of the Optimization Cutoff}
A fundamental question remains: Is the US retreat inevitable, or a consequence of capacity neglect? In Panel B of Figure \ref{fig:sensitivity}, we simulate a counterfactual ``High-Capacity US'' scenario where the US restores its institutional capacity to match China’s current level ($K_{US} = K_{CN} = 8.0$).

\begin{itemize}
    \item \textbf{The Capacity Dividend:} The results are striking. If the US were to close its infrastructure gap ($K=5.0 \rightarrow 8.0$), its optimal openness would surge from $M^* \approx 5.7$ to $M^* \approx 10.5$.
    \item \textbf{Surpassing the Challenger:} Crucially, this counterfactual optimum ($M^* \approx 10.5$) is \textit{higher} than China's optimum ($M^*_{CN} \approx 9.0$), due to the US's higher baseline technological efficiency ($\alpha$).
\end{itemize}

\textbf{Conclusion:} This analysis delivers a pivotal insight. The current crisis is not because the US economy has structurally outgrown the benefits of globalization. On the contrary, it possesses the highest theoretical ceiling. The crisis is the direct result of the historical ``Optimization Cutoff.'' By allowing public capacity to stagnate at 5.0, the US artificially lowered its own ceiling, forcing a rational but costly retreat.
\subsection{The Political Economy of the Capacity Trap: Polarization in the Fractured Metropolis}
The aggregate simulations reveal a critical paradox. Expanding Institutional Capacity ($K$) is demonstrably the \textbf{First-Best} strategy for maximizing US national welfare. However, the observed political reality is a systematic reliance on the \textbf{Second-Best} solution—protectionism (reducing $M$). To reconcile this gap, we must examine the political economy of the ``Fractured Metropolis.''

\subsubsection*{4.5.1 The Historical Origin and Structural Consequence: Elite Exit}
The root of the current political stalemate lies in the ``Optimization Cutoff.'' Globalization incentivized the systematic neglect of domestic public capacity ($K_{Public}$) because global expansion offered higher returns.

This historical path led to a critical structural divergence. While $K_{Public}$ stagnated, the economic beneficiaries of globalization invested in \textbf{Private Capacity ($K_{Private}$)}, which includes exclusive resources such as customized logistics, specialized legal services, and private education. This divergence created two distinct constituencies:
\begin{itemize}
    \item \textbf{The Congested Incumbents:} Reliant on the stagnant $K_{Public}$.
    \item \textbf{The Insulated Elites:} Accessing the efficient $K_{Private}$.
\end{itemize}

Crucially, the expansion of $K_{Private}$ enabled the \textbf{``Elite Exit''} from the public system. When $K_{Public}$ deteriorates, those with the means opt out and switch to private alternatives rather than mobilizing to fix the public system \citep{hirschman1970}.

\subsubsection*{4.5.2 Visualizing the Fractured Metropolis: A Polarization Simulation}
\begin{figure}[H]
    \centering
    \includegraphics[width=\textwidth]{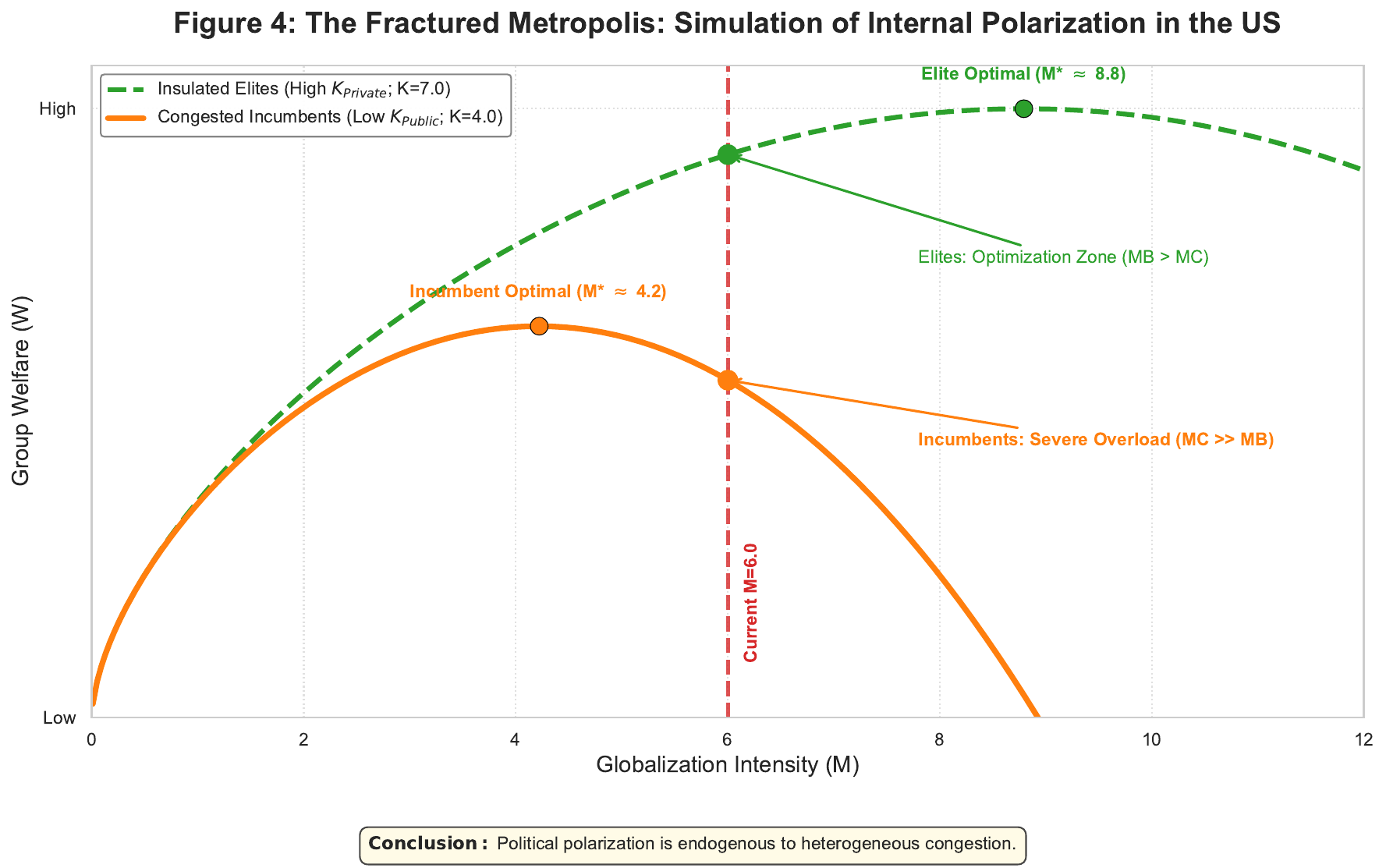} 
    \caption{The Fractured Metropolis: Simulation of Internal Polarization in the US. At current $M=6.0$, Elites (Green) remain in the optimization zone, while Incumbents (Orange) experience severe overload.}
    \label{fig:polarization}
\end{figure}

To illustrate this internal polarization, we simulate the welfare curves for the two heterogeneous groups within the US (Figure \ref{fig:polarization}). We disaggregate the aggregate US parameters to reflect their different realities: Incumbents ($K=4.0$) vs. Elites ($K=7.0$).

The simulation reveals the spatial conflict. At $M=6.0$:
\begin{itemize}
    \item \textbf{Elites:} Still in their optimization zone ($MB > MC$), favoring status quo or openness.
    \item \textbf{Incumbents:} Far to the right of their peak ($MC \gg MB$), experiencing severe welfare losses.
\end{itemize}

\subsubsection*{4.5.3 The Mechanism of Paralysis: The Capacity Trap}
The interaction between Heterogeneous Capacity and the political process creates the self-reinforcing ``Capacity Trap.''

\begin{itemize}
    \item \textbf{The Investment Deadlock:} The fundamental prerequisite for large-scale public investment is political consensus. In the Fractured Metropolis, this consensus collapses due to the Elite Exit. Elites have weak incentives to support costly upgrades for a system they do not use, locking the political system in a ``War of Attrition'' \citep{alesina1991} over who should bear the burden of adjustment.
    
    \item \textbf{The Congestion-Polarization Feedback Loop:} When $M$ exceeds $K_{Public}$, the resulting congestion disproportionately affects Incumbents. This visible ``chaos'' fuels polarization, which further increases the transaction costs of reaching consensus \citep{alesina1999}.
    
    \item \textbf{The Path of Least Resistance:} Paralyzed by the investment deadlock, the system defaults to the path of least resistance. As \citet{tsebelis2002} argues, the US institutional structure creates a critical asymmetry. Increasing $K_{Public}$ requires sustained legislative consensus (overcoming multiple veto players) and is blocked by polarization. In contrast, reducing $M$ (via executive action like tariffs) is feasible, offering a rapid response despite being economically suboptimal.
\end{itemize}

\begin{figure}[htbp]
    \centering
    \includegraphics[width=1.0\textwidth]{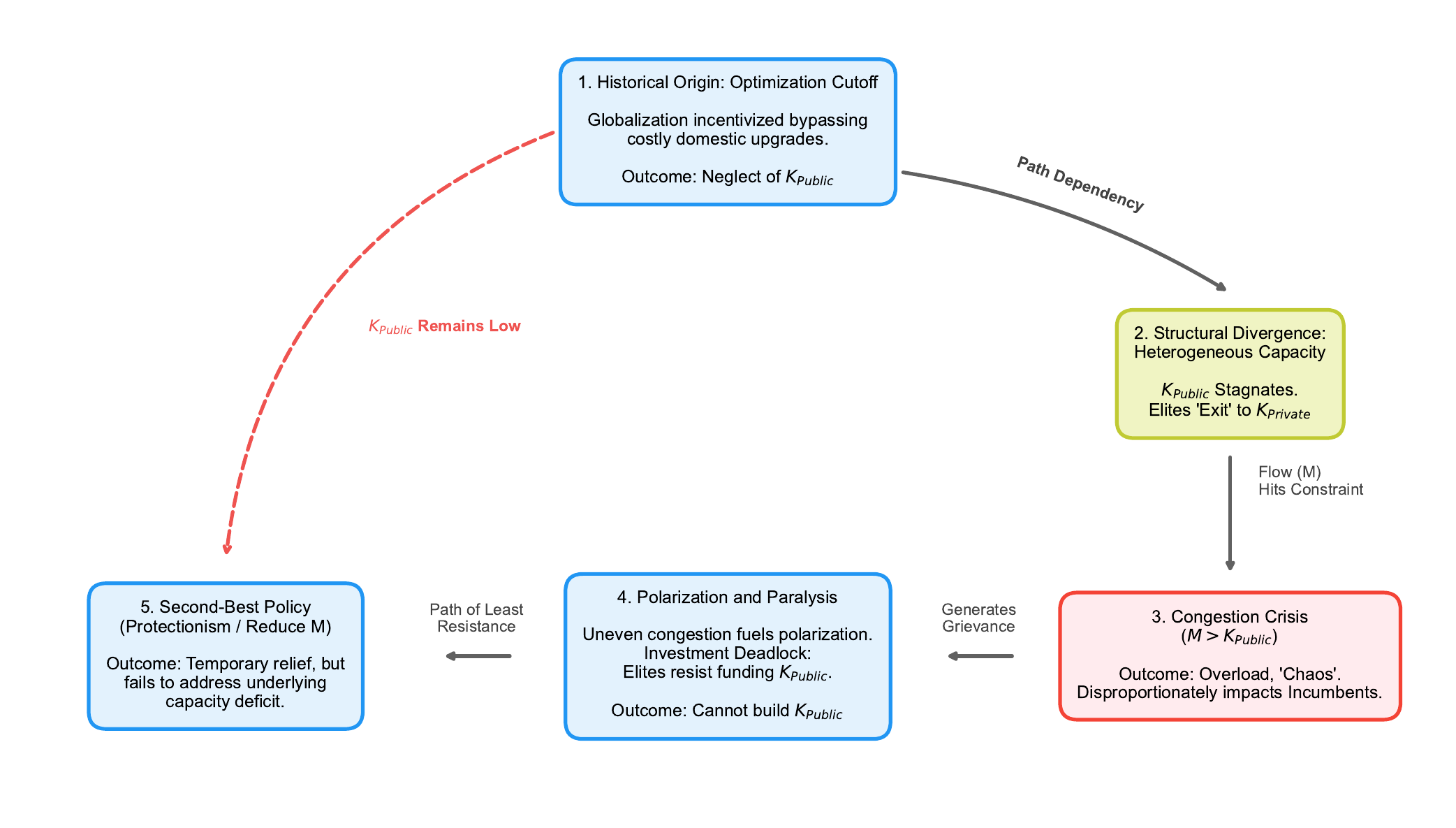} 
    
    \caption{\textbf{The Dynamics of the Capacity Trap in the Fractured Metropolis.} 
    \small This figure illustrates the cycle of neglect where globalization incentivizes the bypassing of domestic public capacity ($K_{Public}$), leading to structural divergence, congestion crisis, and eventually political polarization that locks the system in a low-capacity trap.}
    \label{fig:capacity_trap}
\end{figure}

\subsubsection*{4.5.4 Synthesis}
The ``Capacity Trap'' (Figure \ref{fig:capacity_trap}) is the equilibrium state of the Fractured Metropolis. The polarization between Elites who have exited the public system and Incumbents trapped within it paralyzes the necessary investments in $K_{Public}$, forcing the political system to default to protectionism.
\section{Empirical Evidence: The Collapse of Public Capacity and the Voice of the Incumbents}
Our theoretical framework argues that the ``Great Reversal'' is driven by the stagnation of US Public Institutional Capacity ($K_{Public}$), a legacy of the historical ``Optimization Cutoff.'' This section empirically verifies this causal chain. We demonstrate that the perception of ``chaos'' has surged exponentially and that this perception is tightly synchronized with the material failure of core public institutions.

\subsection{Data Description and Methodology}
To empirically measure the perceived institutional overload, we construct a novel ``Institutional Congestion Index'' using textual analysis of the LexisNexis database. This index tracks the salience of systemic failure associated with globalization.

We focus on the co-occurrence frequency of ``globalization-related terms'' and ``disorder-related terms'' within the United States. To ensure consistency, we restricted our sources to three authoritative pillars of US discourse: \textit{The New York Times} (elite consensus), \textit{The Associated Press} (broad social reality), and \textit{Congressional Documents} (legislative pressure). The search query is defined as: \texttt{(globalization OR "free trade" OR immigration) w/50 (chaos OR crisis OR overwhelmed OR "strain on" OR polarization)}. This ``within 50 words'' constraint ensures that the identified disorder is directly attributed to globalization flows.

\subsection{Trend Analysis: The Political Explosion of the Congested Incumbents}

\begin{figure}[H]
    \centering
    \includegraphics[width=\textwidth]{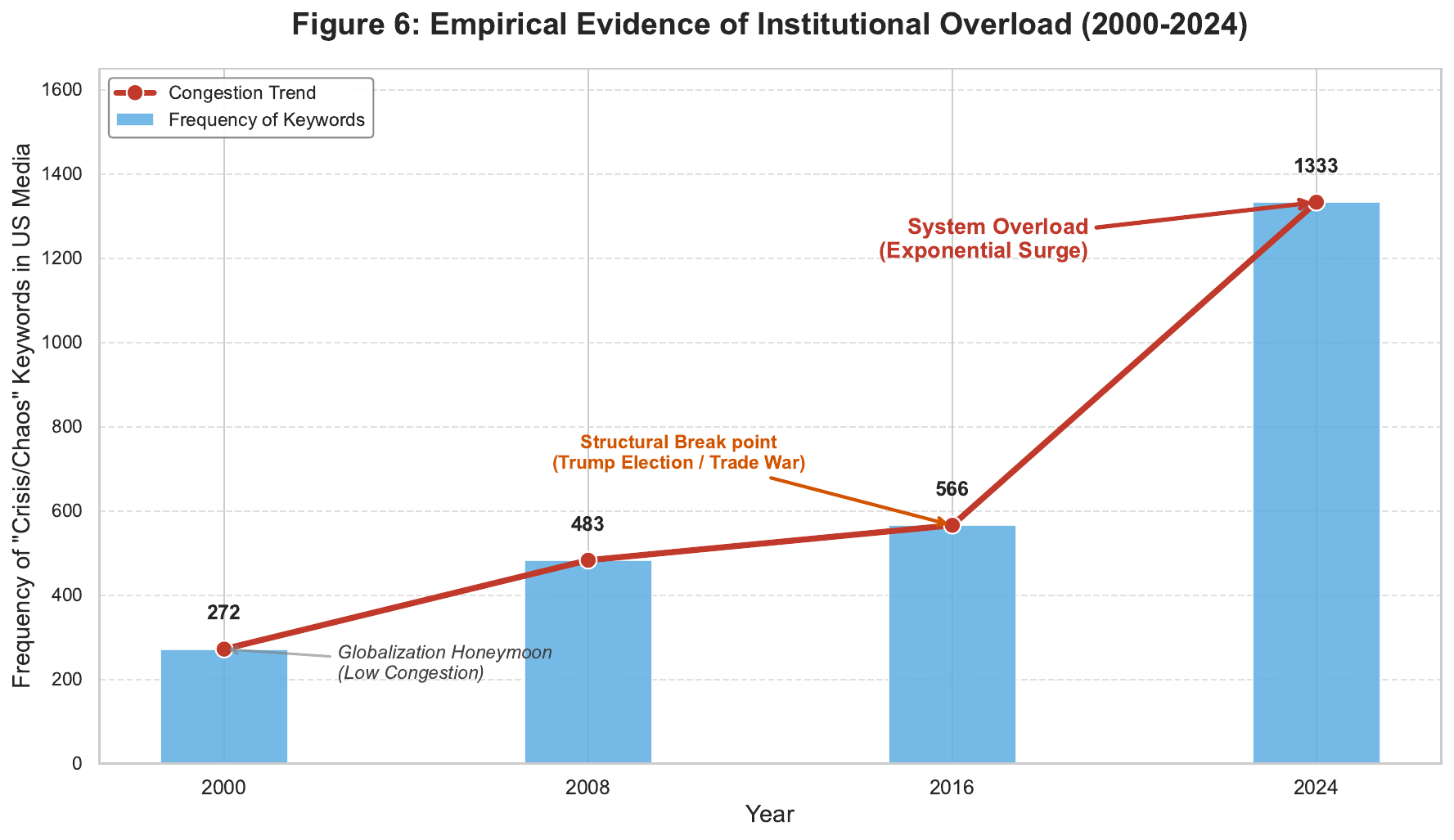} 
    \caption{Empirical Evidence of Institutional Overload (2000-2024). The exponential surge in disorder-related keywords (from 272 to 1,333) signifies the breach of $K_{Public}$ and the resulting political crisis.}
    \label{fig:congestion_index}
\end{figure}

Figure \ref{fig:congestion_index} presents the evolution of the Institutional Congestion Index from 2000 to 2024. The trajectory exhibits a clear non-linear pattern, strongly supporting the convex cost function assumption ($\phi > 1$) in our model. We interpret this trend not merely as rising economic costs, but as the \textit{political manifestation} of the collapse of $K_{Public}$ and the subsequent mobilization of the affected population. The exponential curve reflects the moment when localized congestion escalated into a national political crisis.

The data reveals three distinct phases corresponding to the dynamics of the Optimization Cutoff and the fracturing of the political landscape:

\vspace{0.3cm}
\noindent \textbf{Phase I: The Era of Neglect and Political Silence (2000-2008)}

The congestion index remained relatively low (272 hits in 2000 to 483 in 2008). This corresponds to the peak of the ``Optimization Cutoff.'' Capital was flowing globally, systematically bypassing domestic investment. While $K_{Public}$ began to stagnate due to this neglect, the existing institutional buffer absorbed the initial shocks (e.g., the ``China Shock''). Crucially, the political landscape remained stable because the Elites, insulated by $K_{Private}$, dominated the discourse and maintained a strong consensus favorable to globalization. The Incumbents' grievances remained localized and politically marginalized.

\vspace{0.3cm}
\noindent \textbf{Phase II: The Saturation Point and Erosion of Consensus (2008-2016)}

By 2016, the index reached 566. The cumulative effects of the neglected $K_{Public}$ became increasingly salient as the institutional buffer was exhausted. This period coincides with the rise of populist sentiments, marking a ``structural break.'' The localized congestion experienced by the Incumbents began to systematically erode the elite-driven national political consensus.

\vspace{0.3cm}
\noindent \textbf{Phase III: The Overload and the Voice of the Incumbents (2016-2024)}

The most striking finding is the explosive growth in the recent decade. By 2024, the index skyrocketed to 1,333, a 2.3-fold increase since 2016. This exponential surge signifies that the failure of $K_{Public}$ has breached a critical threshold. We interpret this explosion of ``chaos'' and ``overwhelmed'' discourse as the ``Voice of the Congested Incumbents'' finally breaking through the insulation of the elites. The localized experience of overload has successfully dominated the national political agenda, forcing the entire system into the ``Dis-economy Zone'' and triggering the dynamics of the Capacity Trap.

\subsection{Triangulation: The Material Failure of \texorpdfstring{$K_{Public}$}{K\_Public}}

To validate that our textual index reflects actual systemic failures rather than mere partisan rhetoric, we cross-reference it with ``hard'' indicators of physical and administrative overload. Crucially, these indicators are selected because they represent the failure of core \textit{public} institutional capacities ($K_{Public}$), which our theory identifies as the binding constraint.

\begin{figure}[H]
    \centering
    \includegraphics[width=\textwidth]{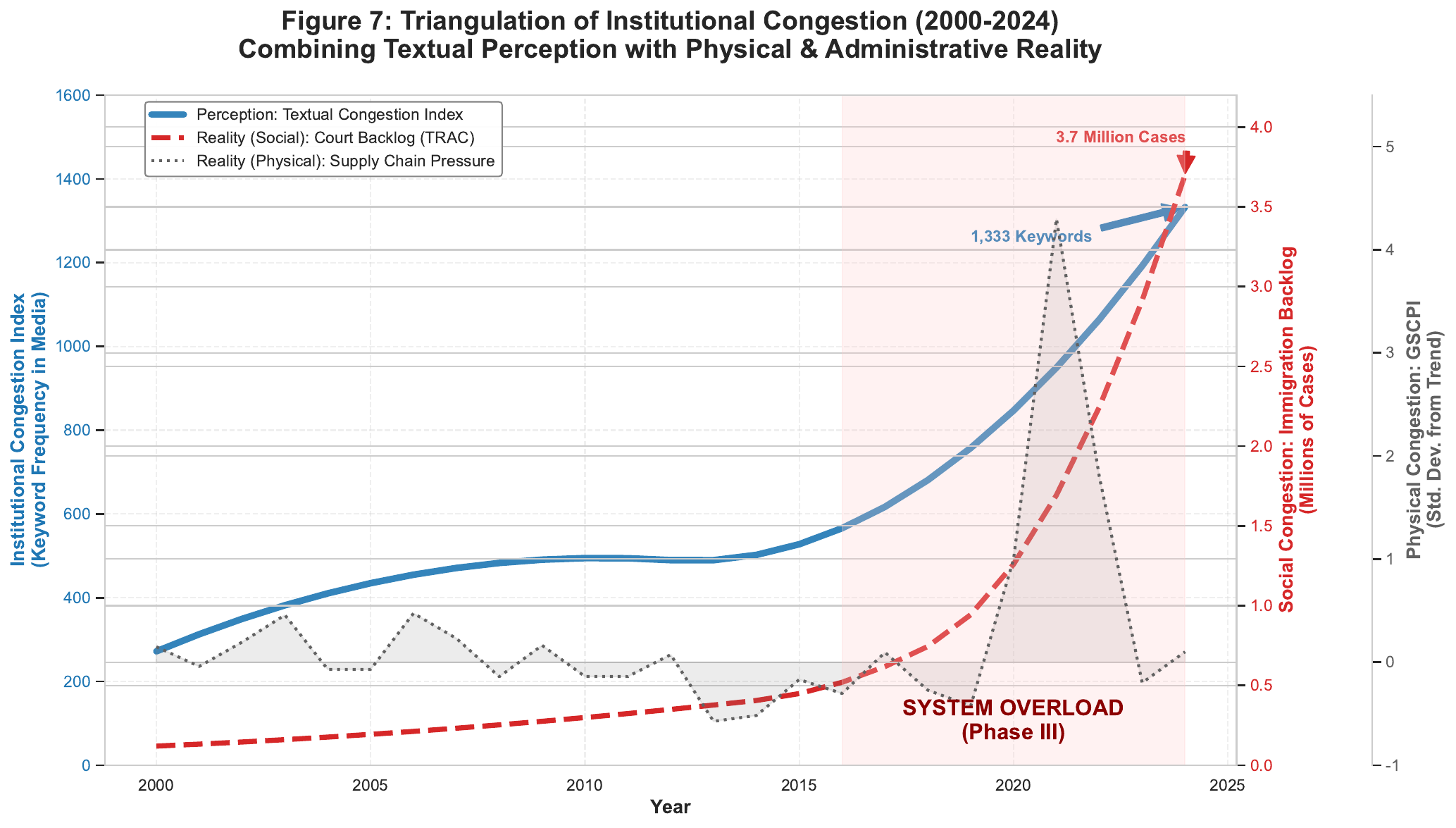} 
    \caption{Triangulation of Institutional Congestion (2000-2024). The perception of chaos (blue line) is tightly synchronized with material reality: the vertical spike in supply chain pressure (grey dotted) and the exponential explosion of immigration court backlogs (red dashed).}
    \label{fig:triangulation}
\end{figure}

\vspace{0.3cm}
\noindent \textbf{1. Administrative Overload: The Judicial Backlog ($K_{Gov}$) – The Fractured Metropolis in Action}

We utilize data from the Transactional Records Access Clearinghouse (TRAC, 2024) regarding US Immigration Court backlogs. This is a definitive proxy for the failure of public administrative capacity to scale with globalization flows ($M$).

\textbf{The Trend:} The data reveals a strictly exponential trajectory mirroring our textual index. Pending cases rose from approximately 500,000 in 2015 to an unprecedented \textbf{3.7 million by 2024}.

\textbf{The Interpretation (The Fractured Metropolis):} This 7-fold increase represents a material collapse of $K_{Public}$. This failure is politically explosive because the resulting congestion costs are distributed heterogeneously. The administrative paralysis generates immense localized chaos (social friction, overwhelmed local services) that disproportionately affects the \textit{Congested Incumbents} (e.g., border communities, taxpayers funding public services). Simultaneously, the \textit{Insulated Elites}, relying on specialized legal services and networks ($K_{Private}$), can largely bypass this chaos. This divergence validates the polarization mechanism central to our framework.

\vspace{0.3cm}
\noindent \textbf{2. Physical Overload: The Supply Chain Spike ($K_{Phy}$) – The Legacy of the Optimization Cutoff}

We examine the Global Supply Chain Pressure Index (GSCPI) constructed by the Federal Reserve Bank of New York (2024).

\textbf{The Spike:} The GSCPI remained stable until 2019 but exhibited a vertical surge, peaking at \textbf{+4.3 standard deviations} in 2021-2022.

\textbf{The Interpretation (The Optimization Cutoff):} This physical breakdown of logistics infrastructure (ports, roads) is not merely a reaction to an exogenous shock; it is the direct consequence of the long-term underinvestment identified as the \textit{Optimization Cutoff} (Section 4.1). The historical neglect of $K_{Public}$—driven by the prioritization of global expansion over domestic upgrades—left the physical system brittle and unable to absorb the flow shock. The crisis is the manifestation of this historical path dependency.

\vspace{0.3cm}
\noindent \textbf{3. Synthesis: From Perception to Reality}

Figure \ref{fig:triangulation} demonstrates the tight synchronization between the ``soft'' perception (the Voice of the Incumbents) and the ``hard'' reality (the material failure of $K_{Public}$). It provides robust evidence that the US public system has breached its critical capacity threshold ($M > K_{Public}$), validating the core mechanism of our model and confirming that the political crisis is rooted in material institutional failure.

\subsection{Causal Mechanism Illustration: The 2021-2022 Supply Chain Crisis and Heterogeneous Impact}

The 2021-2022 US West Coast port congestion crisis provides a vivid demonstration of our theoretical framework, illustrating how the failure of $K_{Public}$ generates exponential costs and how $K_{Private}$ insulates the elites.

\paragraph{The Root Cause: The Legacy of the Optimization Cutoff}
The crisis was not caused by the demand shock itself, but by the pre-existing fragility of $K_{Public}$. Decades of underinvestment (the Optimization Cutoff) meant that critical infrastructure like the Ports of Los Angeles and Long Beach were operating near saturation \cite{portla2022}. The system lacked the physical buffer (dock space, chassis availability, automation) to absorb shocks \cite{asce2021, wbsp2022}.

\paragraph{The Overload ($M > K_{Public}$)}
The pandemic-driven surge in imports collided directly with this stagnant capacity ceiling. As $M$ exceeded $K_{Public}$, the system gridlocked. A record 109 container ships were waiting offshore in October 2021 \cite{meyers2021}. The price of shipping a container from China to the US West Coast skyrocketed from under \$2,000 to over \$20,000 \cite{freightos2021}.

\paragraph{The Heterogeneous Impact: The Fractured Metropolis in Crisis}
Crucially, the costs of this congestion were not borne equally. The crisis highlighted the sharp divergence between $K_{Public}$ and $K_{Private}$:

\begin{itemize}
    \item \textbf{The Insulated Elites (High $K_{Private}$):} Large multinational corporations demonstrated significant $K_{Private}$. Major retailers (e.g., Walmart, Amazon, Home Depot) mitigated the impact by chartering their own ships, utilizing alternative, less congested ports, and leveraging their market power to secure priority access \cite{whitehouse2022}. They effectively bypassed the most severe bottlenecks in the public system.
    
    \item \textbf{The Congested Incumbents (SMEs and Consumers):} Small and medium-sized enterprises (SMEs), heavily reliant on the public logistics network, lacked the resources to bypass the gridlock. They faced crippling delays, exponential cost increases, and inventory shortages that threatened their solvency. Consumers faced inflation and product unavailability. They bore the brunt of the chaos.
\end{itemize}

This divergence in outcomes explains the political dynamic. The systemic failure of $K_{Public}$ generated widespread economic pain among the Incumbents, further fueling the political demand for ``reshoring'' and protectionism (reducing $M$) as a response to the unreliability of the globalized system.

\subsection{Robustness Check: A Placebo Test}
To verify that the exponential surge in our Index reflects a specific systemic failure rather than a general inflation of ``crisis discourse,'' we conduct a placebo test using a control topic: \textbf{K-12 Education}.

\begin{figure}[H]
    \centering
    \includegraphics[width=\textwidth]{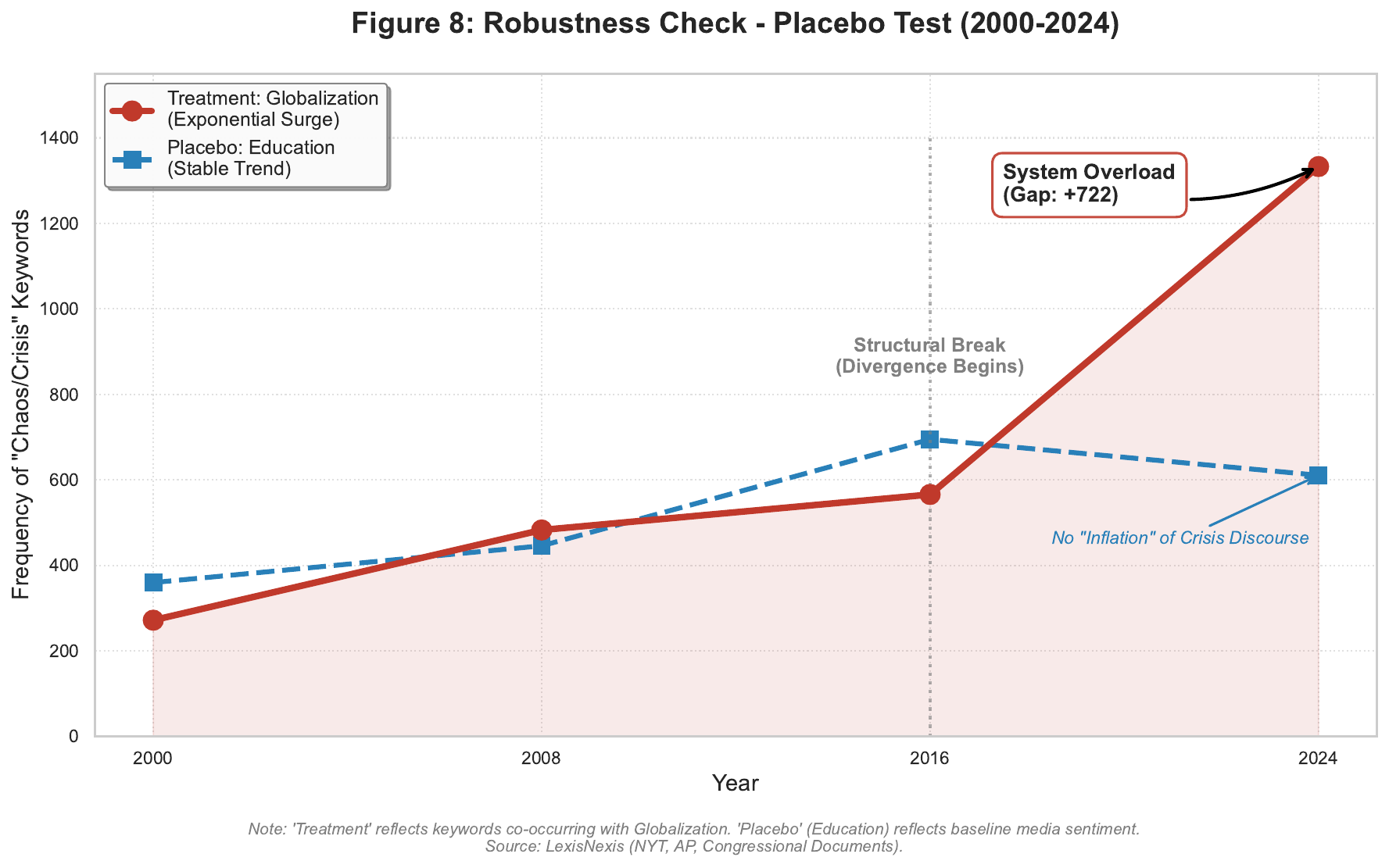}
    \caption{Robustness Check: Placebo Test (2000-2024). While the ``chaos'' trend for Education remained stable, the Globalization trend surged exponentially, confirming a specific systemic overload.}
    \label{fig:placebo_test}
\end{figure}

As shown in Figure \ref{fig:placebo_test}, while the ``congestion'' sentiment for Education remained relatively stable between 2016 (695 hits) and 2024 (611 hits), the Globalization index skyrocketed from 566 to 1,333 during the same period. This divergence confirms that the observed ``overload'' is not a linguistic artifact, but a specific symptom of external flows exceeding institutional capacity ($K_{Public}$).

\subsection{Source Heterogeneity Analysis: The Escalation Pathway}
Analyzing the temporal dynamics across our three sources (NYT, AP, Congressional Documents) allows us to trace the escalation pathway of the congestion crisis—from social perception to legislative action.

\paragraph{The Tipping Point (Post-2016):} The exponential surge in 2024 is disproportionately driven by an explosion in \textbf{Congressional Documents} (e.g., committee hearings, proposed bills).

\paragraph{Interpretation:} This trajectory reveals the mechanism of political response: \textit{Societal Impact $\rightarrow$ Media Attention $\rightarrow$ Legislative Action}. The overwhelming dominance of legislative records signifies that the grievances of the Congested Incumbents have escalated from a manageable social issue to an acute \textbf{Governance Crisis}. It confirms that the political system is now fully mobilized in response to the overload, leading to the dynamics of the ``Capacity Trap'' (Section 4.5) where protectionism becomes the path of least resistance.
\section{Conclusion and Policy Implications}

\subsection{Summary of Findings: The Anatomy of the Fractured Metropolis}
This paper revisits the paradox of the ``Great Reversal'' of globalization. By conceptualizing the national economy as a ``Congestible Club Good,'' we identify Institutional Capacity ($K$)—rather than ideology—as the binding constraint.

Theoretically, we contribute three key concepts:
\begin{itemize}
    \item \textbf{The Optimization Cutoff:} We provide an endogenous explanation for US capacity stagnation. Globalization incentivized bypassing domestic upgrades, leading to a path dependency of neglect.
    \item \textbf{Heterogeneous Capacity and Polarization:} We demonstrate that political conflict is endogenous to uneven congestion. The divergence between stagnant $K_{Public}$ and efficient $K_{Private}$ creates the split between ``Congested Incumbents'' and ``Insulated Elites.''
    \item \textbf{The Capacity Trap:} We explain the default to protectionism. The ``Elite Exit'' paralyzes the consensus required to fund $K_{Public}$, locking the system into a suboptimal equilibrium.
\end{itemize}

Empirically, our textual Institutional Congestion Index and triangulation with physical data (GSCPI, Immigration Backlogs) confirm that the US public system has breached its critical capacity threshold.

\subsection{Policy Implications: Escaping the Trap}
Our findings suggest that the current trajectory of the global economy is driven by a deep structural asymmetry between the US (the saturated ``inner city'') and China (the expansive ``suburb''). This asymmetry requires distinct policy responses.

\subsubsection*{For the United States: The Political Challenge of Rebuilding $K_{Public}$}
The standard policy prescription—``increase infrastructure investment''—is insufficient because it ignores the political deadlock identified in the Capacity Trap. The core challenge for the US is not merely identifying the need for $K_{Public}$, but overcoming the political paralysis caused by the ``Elite Exit.''

\begin{itemize}
    \item \textbf{Beyond Protectionism:} The current reliance on tariffs and border walls (reducing $M$) is a Second-Best solution. While it may temporarily alleviate the symptoms of congestion for the Incumbents, it sacrifices the long-term economic gains of globalization and fails to address the underlying structural fragility.
    \item \textbf{Breaking the Deadlock:} Escaping the Capacity Trap requires a fundamental political realignment that rebuilds the consensus for funding public goods. This involves addressing the structural incentives that encourage the Elite Exit and finding mechanisms to finance the massive investments needed to close the maintenance deficit \citep{melusen2025, asce2021}.
\end{itemize}

Unless this political deadlock is broken, protectionism will likely remain the long-term equilibrium state for the US. The priority must shift from ``building walls'' (flow restriction) to ``building courts and ports'' (capacity expansion), but this requires a political, not just technical, solution.

\subsubsection*{For China: Managing the Dynamics of Expansion}
China currently benefits from a ``Capacity Surplus'' ($K_{CN} > M_{CN}$) and the catch-up effect ($\delta$), allowing it to rationally support global openness. However, the US experience offers a cautionary lesson.

\begin{itemize}
    \item \textbf{The Limits of the Catch-up Effect:} As China approaches the technological frontier, $\delta$ will diminish, and the marginal costs of domestic optimization will rise.
    \item \textbf{Avoiding the US Trap:} To sustain its integration, China must proactively manage its capacity. Continued investment in ``hard'' infrastructure must be matched by upgrades in ``soft'' institutional capacity (legal frameworks, social safety nets) to prevent the onset of exponential congestion costs and the emergence of its own ``Optimization Cutoff'' dynamics in the future.
\end{itemize}

\subsubsection*{For Global Governance: From Flow to Capacity}
The ``Fractured Metropolis'' framework suggests a fundamental shift in the focus of global governance. The 20th-century focus on reducing barriers to flow (e.g., tariffs) is no longer sufficient. The 21st-century challenge is managing capacity. Just as a metropolis cannot remain open without upgrading its subway system, the global liberal order cannot survive without rebuilding the domestic institutional capacity of its core economies.

\subsection{Limitations and Future Research}

This study acknowledges certain limitations, primarily concerning the measurement and identification of Institutional Capacity ($K$).

\vspace{0.3cm}
\noindent \textbf{1. Measurement Challenges of Heterogeneous Capacity}

While our model parameters are calibrated based on authoritative stylized facts, they remain inputs for simulation rather than structural estimation. A significant empirical challenge lies in precisely quantifying $K$ and, more critically, empirically distinguishing between Public Capacity ($K_{Public}$) and Private Capacity ($K_{Private}$). The ``Fractured Metropolis'' mechanism hinges on this divergence, yet data on the utilization of $K_{Private}$ is often difficult to quantify systematically using aggregate data.

\vspace{0.3cm}
\noindent \textbf{2. Avenues for Future Research}

To address these limitations and strengthen the causal identification of our proposed mechanisms, future research should focus on developing more granular empirical strategies.

\begin{itemize}
    \item \textbf{Micro-Level Analysis of $K_{Private}$ (Elite Exit):} A promising avenue involves utilizing firm-level or individual-level microdata to directly measure the utilization of Private Capacity and identify the ``Elite Exit.'' Analyzing expenditures on resources that substitute for public infrastructure—such as customized logistics, specialized legal/financial services, private security, or elite private education—could provide a direct proxy for $K_{Private}$. Testing whether firms or individuals investing heavily in $K_{Private}$ exhibit different political preferences regarding globalization would directly validate the ``Insulated Elite'' mechanism.
    
    \item \textbf{Regional Panel Data Analysis:} Our empirical triangulation relies on aggregate national data. Utilizing regional-level panel data could provide a more granular understanding of how localized congestion drives polarization. By examining the correlation between state-level infrastructure quality (local $K_{Public}$), exposure to globalization shocks ($M$), and voting patterns, researchers can better isolate the impact of capacity constraints on political outcomes and test the Congested Incumbents hypothesis.
\end{itemize}

Nevertheless, by introducing the concepts of the Optimization Cutoff and Heterogeneous Capacity, this paper provides a novel analytical framework for understanding why the world's architect of openness has become its foremost skeptic, and why its society is so deeply divided over the path forward.

\section*{Acknowledgements}
The conceptualization, theoretical framework, and core arguments presented in this paper are the original work of the author. The author acknowledges the assistance of Google's Gemini for linguistic refinement and for generating the Python scripts used for data visualization. All final content, analysis, and errors remain the sole responsibility of the author.


\appendix
\section{Appendix A: Mathematical Derivations of the Equilibrium Model}

This appendix provides the formal mathematical derivations for the theoretical framework presented in Section 3 and the simulations in Section 4. It details the welfare maximization problem, the equilibrium solution, comparative statics, and the formalization of the heterogeneous capacity extension.

\subsection{A.1. Model Setup}

The objective is to maximize the Net National Welfare ($W$), defined as the difference between the economic benefits of globalization ($B(M)$) and the social congestion costs ($C(M)$).

\paragraph{The Benefit Function:}
\begin{equation}
    B(M) = \alpha (1 + \delta) M^\theta
\end{equation}
Where $\alpha > 0$ (Baseline technology), $0 < \theta < 1$ (Elasticity of benefits, ensuring diminishing returns), and $\delta \ge 0$ (Catch-up Parameter).

\paragraph{The Congestion Cost Function:}
\begin{equation}
    C(M) = \gamma \left( \frac{M}{K} \right)^\phi = \gamma K^{-\phi} M^\phi
\end{equation}
Where $\gamma > 0$ (Sensitivity to disorder), $\phi > 1$ (Congestion Elasticity, ensuring convexity/exponential costs), and $K > 0$ (Institutional Capacity).

\subsection{A.2. Welfare Optimization and Equilibrium Solution}

The maximization problem is:
\begin{equation}
    \max_{M} W(M) = \alpha (1 + \delta) M^\theta - \gamma K^{-\phi} M^\phi
\end{equation}

The First Order Condition (FOC) requires Marginal Benefit (MB) equals Marginal Congestion Cost (MC):
\begin{align}
    MB(M) &= \frac{dB}{dM} = \alpha (1 + \delta) \theta M^{\theta - 1} \\
    MC(M) &= \frac{dC}{dM} = \gamma \phi K^{-\phi} M^{\phi - 1}
\end{align}

Setting $MB(M) = MC(M)$:
\begin{equation}
    \alpha (1 + \delta) \theta M^{\theta - 1} = \gamma \phi K^{-\phi} M^{\phi - 1}
\end{equation}

Solving for $M$ yields the optimal globalization level $M^*$. Rearranging the terms:
\begin{equation}
    M^{\phi - \theta} = K^{\phi} \cdot \left[ \frac{\alpha (1 + \delta) \theta}{\gamma \phi} \right]
\end{equation}

The explicit solution for $M^*$ is (Note: Since $\phi > 1$ and $\theta < 1$, implies $\phi - \theta > 0$):
\begin{equation} \label{eq:optimal_M}
    M^* = K^{\frac{\phi}{\phi - \theta}} \cdot \underbrace{\left[ \frac{\alpha (1 + \delta) \theta}{\gamma \phi} \right]^{\frac{1}{\phi - \theta}}}_{\text{Constant Terms}}
\end{equation}

\subsection{A.3. Second Order Condition (SOC) Verification}

To ensure $M^*$ is a maximum, we require $\frac{d^2W}{dM^2} < 0$.

\begin{equation}
    \frac{d^2W}{dM^2} = \frac{d(MB)}{dM} - \frac{d(MC)}{dM}
\end{equation}

Analyzing the slopes:
\begin{itemize}
    \item Slope of MB: $\frac{d(MB)}{dM} < 0$ (because $\theta - 1 < 0$).
    \item Slope of MC: $\frac{d(MC)}{dM} > 0$ (because $\phi - 1 > 0$).
\end{itemize}

Therefore:
\begin{equation}
    \frac{d^2W}{dM^2} = (\text{Negative Value}) - (\text{Positive Value}) < 0
\end{equation}

The SOC is globally satisfied, confirming $M^*$ is the unique maximum.
\subsection{A.4. Comparative Statics}

We analyze how $M^*$ responds to shifts in key parameters.

\paragraph{The Impact of Institutional Capacity ($K$) (Proposition 1)}
We examine $\frac{\partial M^*}{\partial K}$. From Equation (\ref{eq:optimal_M}), since the exponent $\frac{\phi}{\phi - \theta}$ is strictly positive (given $\phi > 1, \theta < 1$):
\begin{equation}
    \frac{\partial M^*}{\partial K} > 0
\end{equation}
\textbf{Result:} The optimal level of globalization is strictly increasing in institutional capacity.

\paragraph{The Impact of the Catch-up Parameter ($\delta$) (Proposition 1)}
We examine $\frac{\partial M^*}{\partial \delta}$. $M^*$ is proportional to $(1 + \delta)^{\frac{1}{\phi - \theta}}$. Since the exponent is positive:
\begin{equation}
    \frac{\partial M^*}{\partial \delta} > 0
\end{equation}
\textbf{Result:} Economies with higher learning potential (high $\delta$) have a higher optimal openness threshold.

\subsection{A.5. Formalizing the Heterogeneous Capacity Model (The Fractured Metropolis)}

To mathematically ground the political polarization analyzed in Section 4.5, we extend the framework by disaggregating the population into two groups: Insulated Elites ($E$) and Congested Incumbents ($I$).

\paragraph{Assumptions:}
\begin{enumerate}
    \item \textbf{Common Exposure:} Both groups are exposed to the same level of aggregate globalization intensity ($M$).
    \item \textbf{Heterogeneous Capacity:} The groups differ in the institutional capacity they access. Elites access high Private Capacity ($K_E$), while Incumbents rely on low Public Capacity ($K_I$).
\end{enumerate}

The core assumption of the Fractured Metropolis is:
\begin{equation}
    K_E > K_I
\end{equation}

\paragraph{Group Welfare Functions:}
The welfare function for group $j \in \{E, I\}$ is (assuming similar benefit parameters to isolate the effect of capacity):
\begin{equation}
    W_j(M) = B(M) - C_j(M) = \alpha (1 + \delta) M^\theta - \gamma \left( \frac{M}{K_j} \right)^\phi
\end{equation}

\paragraph{Divergence in Optimization and the Root of Polarization:}
We apply the equilibrium condition derived in A.2 to each group to find their respective preferred globalization levels ($M^*_j$):
\begin{equation}
    M^*_j = K_j^{\frac{\phi}{\phi - \theta}} \cdot \left[ \text{Constant Terms} \right]
\end{equation}

From the comparative static result in A.4 ($\frac{\partial M^*}{\partial K} > 0$), it mathematically follows that:
\begin{equation}
    \text{If } K_E > K_I \implies M^*_E > M^*_I
\end{equation}
This formally proves that the group with higher capacity rationally prefers a higher level of globalization.

\paragraph{Political Conflict at Equilibrium:}
The political conflict described in the Capacity Trap arises when the actual globalization level ($M_{Actual}$) falls between the two preferred optima:
\begin{equation}
    M^*_I < M_{Actual} < M^*_E
\end{equation}

In this scenario:
\begin{itemize}
    \item \textbf{Incumbents} are in their ``Dis-economy Zone'' ($M_{Actual} > M^*_I$) and demand protectionism.
    \item \textbf{Elites} remain in their ``Optimization Zone'' ($M_{Actual} < M^*_E$) and support openness.
\end{itemize}

This formalizes the central argument that political polarization is endogenous to the heterogeneous distribution of institutional capacity.
\end{document}